\documentclass[10pt]{iopart}
\usepackage{amssymb}

\begin{document}

\title{Self-similar Bianchi models: II. Class B models}
\author{Pantelis S. Apostolopoulos$^{\dag }$}

\begin{abstract}
In a companion article (referred hearafter as paper I) a detailed study of
the simply transitive Spatially Homogeneous (SH) models of class A
concerning the existence of a simply transitive similarity group has been
given. The present work (paper II) continues and completes the above study
by considering the remaining set of class B models. Following the procedure
of paper I we find all SH models of class B subjected only to the minimal geometric 
assumption to admit a proper Homothetic Vector Field (HVF). The physical 
implications of the obtained geometric results are studied by specialising our 
considerations to the case of vacuum and $\gamma -$law perfect fluid 
models. As a result we regain all the known exact solutions regarding vacuum
and non-tilted perfect fluid models. In the case of tilted fluids we find
the \emph{general }self-similar solution for the exceptional type VI$_{-1/9}$
model and we identify it as equilibrium point in the corresponding dynamical
state space. It is found that this \emph{new} exact solution belongs to the
subclass of models $n_\alpha ^\alpha =0$, is defined for $\gamma \in (\frac
43,\frac 32)$ and although has a five dimensional stable manifold there
exist always two unstable modes in the restricted state space. Furthermore
the analysis of the remaining types, guarantees that tilted perfect fluid
models of types III, IV, V and VII$_h$ cannot admit a proper HVF strongly
suggesting that these models either may not be asymptotically self-similar
(type V) or may be extreme tilted at late times. Finally for each Bianchi
type, we give the extreme tilted equilibrium points of their state space.
\end{abstract}

\address{\dag University of Athens, Department of Physics, Nuclear and
Particle Physics Section, Panepistimiopolis, Zografos 15771, Athens, Greece.}
\ead{papost@phys.uoa.gr\newline
} \relax
\maketitle

\section{Introduction}

\setcounter{equation}{0}

In a previous article \cite{Apostol9} (designated in the following as paper
I) we presented a general study of the Spatially Homogeneous (SH) models of
class A admitting a four dimensional group of homotheties acting simply
transitively on the spacetime manifold. We concentrated on tilted $\gamma-$%
law perfect fluids and we have given various results concerning the
existence of an admissible self-similar perfect fluid model in all the
Bianchi types of class A. As a consequence this analysis has led us to
regain the type II solution \cite{Hewitt}, to show the non-existence of
self-similar type VII$_0$,VIII and IX tilted perfect fluid models \cite{Apostol9,Apostol-Tsampa7} 
and to prove the generality of the known perfect fluid solution for type VI$_0$ models \cite
{Apostol12,Rosquist-Jantzen1}.

Our intention in the present work is to address the problem regarding the
existence of a proper Homothetic Vector Field (HVF) for the remaining SH
models namely those of class B and to examine the physical and dynamical
implications of the geometric results with the hope to provide a way of
obtaining a better understanding of their past and future asymptotic
dynamics.

Section 2 contains some basic elements of the two major frameworks of
studying SH cosmologies: the \emph{orthonormal frame} and \emph{metric}
approaches. In particular and adopting the conventions and the methodology
of \cite{Hewitt-Bridson-Wainwright}, we briefly discuss the structure of the
dynamical state space of the SH models, coming from the evolution equations
and the algebraic restrictions for the case where the spacetime is filled
with a tilted (in general) $\gamma-$law perfect fluid. In section 3, by exploiting 
the basic relations of the metric approach and the results of paper I 
we find all the (simply) transitively self-similar SH models i.e. those models which 
admit a proper Homothetic Vector Field (HVF) without assuming a specific matter 
content filled the spacetime. From these results we proceed in section 4 and give an 
analysis of their possible physical interpretation. Due to their sound dynamical significance, 
we apply the general geometric results only to the case of vacuum and $\gamma-$law 
perfect fluid models. For illustration purposes we reproduce the known vacuum and 
non-tilted self-similar models of type III, IV, VI$_h$. In the case of tilted perfect fluids 
we find a \emph{new exact solution}, representing the \emph{general} self-similar tilted 
model of type VI$_{-1/9}$. Finally section 5 contains a summary of all the known 
self-similar tilted perfect fluid solutions and a brief discussion regarding
their importance in the asymptotic behaviour of generic models.

Throughout the following conventions have been used: spatial frame indices
are denoted by lower Greek letters $\alpha ,\beta ,...=1,2,3$, lower Latin
letters denote spacetime indices $a,b,...=0,1,2,3$ and we use geometrised
units such that $8\pi G=c=1$.

\section{Preliminaries}

\setcounter{equation}{0}

Using the so-called \emph{metric }approach, one can easily derive the
explicit form (in local coordinates) of the self-similar metric and the
conformally mapped tilted fluid velocity. Due to the fact that the reduced
Field Equations (FE) have purely algebraic form, the determination of a
specific self-similar vacuum or $\gamma-$law perfect fluid model is 
straightforward. On the other hand the \emph{orthonormal frame} approach 
is used to write the FE in an autonomous form and the resulting system of 
decoupled first order differential equations is studied in the \emph{dynamical 
state space} of the corresponding model \cite{Wainwright-Ellis}. Since 
every (non-extreme) equilibrium point is represented by an exact self-similar 
solution the mutual use of the approaches is able to show the existence or not 
of a self-similar model and to provide with the necessary tools for the description 
of their future asymptotic dynamics. Therefore and in order the present article to be 
self-contained as possible we find convenient to give the general setup
concerning the essence of both approaches which shall be used frequently in
what follows (the reader is referred to \cite
{Hewitt-Bridson-Wainwright,Wainwright-Ellis,Ellis-MacCallum,MacCallum2,Ryan-Shepley1} for
an extensive exposition of the formalism used in this section).

\subsection{The metric approach}

In Bianchi models the existence of a $\mathcal{G}_3$ Lie algebra of Killing
Vector Fields (KVFs) $\mathbf{X}_\alpha $ with 3-dimensional spacelike
orbits $\mathcal{C}$ implies the existence of a uniquely defined unit
timelike congruence $n^a$ ($n^an_a=-1$) normal to the spatial foliations $%
\mathcal{C}$:

\begin{equation}
n_{[a;b]}=0=n_{a;b}n^b\Leftrightarrow \frac 12\mathcal{L}_{\mathbf{n}%
}g_{ab}\equiv n_{a;b}=\sigma _{ab}+\frac \theta 3h_{ab}  \label{prel1}
\end{equation}
where $\sigma _{ab},\theta =n_{a;b}g^{ab},h_{ab}=g_{ab}+n_an_b$ are the
kinematical quantities associated with the $n^a$ according to the standard
kinematical decomposition of an arbitrary timelike congruence \cite{Ellis1}.
Because $n^a$ is irrotational and geodesic, there exists a time function $%
t(x^a)$ such that $n^a=\delta _t^a$ i.e. each value of $t$ essentially
represents the hypersurfaces $\mathcal{C}$. Therefore local coordinates $%
\{t,x^\alpha \}$ can be found in which the line element takes the following
form: 
\begin{equation}
ds^2=-dt^2+g_{\alpha \beta }(t)\omega _a^\alpha \omega _b^\beta dx^adx^b
\label{prel2}
\end{equation}
where we have employed a basis of invariant vector fields and $1-$forms constituting of $%
\mathbf{e}_\alpha $ and its dual $\mathbf{\omega }^\beta $.

Furthermore if we assume that the spacetime is filled with a perfect fluid
having a linear equation of state $\tilde{p}=(\gamma -1)\tilde{\mu}$, the
Ricci tensor is written:

\begin{equation}
R_{ab}=\gamma \tilde{\mu}u_au_b+\frac{(2-\gamma )\tilde{\mu}}2g_{ab}
\label{prel4}
\end{equation}
where $\tilde{\mu},\tilde{p}$ are the energy density and the pressure
measured by the observers comoving with the fluid velocity $u^a$. The
tilted (in general) fluid velocity $u^a$ can be decomposed parallel and
normal to $n^a$ as follows:

\begin{equation}
\mathbf{u}=\Gamma \mathbf{n}+\Gamma \Delta _\alpha \mathbf{\omega }^\alpha
=\Gamma \mathbf{n+}\Gamma B^\alpha \mathbf{e}_\alpha  \label{prel5}
\end{equation}
where $B^\alpha (t),\Delta _\alpha (t)$ are the frame components of the spatial part of the 
four-velocity $u^a$ and $\Gamma $ is a smooth function of the time
coordinate $t$ satisfying the constraint: 
\begin{equation}
\Gamma =\left( 1-B^\alpha \Delta _\alpha \right) ^{-\frac 12}  \label{prel6}
\end{equation}
and $B^\alpha \Delta _\alpha <1$.

Finally we note that under the assumption of a $\mathcal{H}_4$ Lie algebra
of homotheties, it has been shown in paper I that $\Gamma $ becomes a
constant and both the frame components of the metric and the fluid velocity
are determined explicitly up to integration constants.

\subsection{The orthonormal frame approach}

Of particular importance in the study of the asymptotic dynamics of SH
models, is the reformulation of the FE as an autonomous
system of first order ordinary differential equations. This has been done in 
\cite{Hewitt-Bridson-Wainwright} where, using the orthonormal frame
approach, the FE are written in terms of a set of expansion-normalised
variables, by using the dimensionless time parameter: 
\begin{equation}
\frac{dt}{d\tau }=\frac 1H,\quad \frac{dH}{d\tau }=-\left( 1+q\right) H
\label{diffequat2}
\end{equation}
where $q,H$ being the deceleration and expansion (Hubble) parameter
respectively of the normal timelike congruence $n^a$.

The resulting system consists of the evolution equations for the components $%
\Sigma_{\alpha\beta}$ of the shear tensor of the unit normal vector field $%
n^a$, the spatial curvature $A_\alpha,n_{\alpha \beta }$ of the orbits of the $G_3$
isometry group (due to Jacobi identities) and the spatial components of the
tilted fluid velocity $v^\alpha$.

Using the freedom of a time-dependent spatial rotation, we may choose the
orthonormal tetrad to be the eigenframe of $n_{\alpha \beta }$ therefore the
contracted form of Jacobi identities $n_{\alpha \beta }A^\beta =0$ implies:

\begin{equation}
n_{\alpha \beta }=\left( 
\begin{array}{lll}
0 & 0 & 0 \\ 
0 & N_2 & 0 \\ 
0 & 0 & N_3
\end{array}
\right) ,\qquad A_\alpha =A_1\delta _\alpha ^1.  \label{def1}
\end{equation}
We note that in types VI$_h$ and VII$_h$ the evolution equations for $%
n_{\alpha \beta }$ and $A^\alpha$ can be used to express the component $A_1$
in terms of the curvature variables: 
\begin{equation}
A_1^2=hN_2N_3.  \label{first_integral}
\end{equation}
In addition and following \cite{Hewitt-Bridson-Wainwright} we introduce the
shear variables: 
\begin{equation}
\Sigma _{+}=\frac 12\left( \Sigma _{22}+\Sigma _{33}\right) ,\quad \Sigma
_{-}=\frac 1{2\sqrt{3}}\left( \Sigma _{22}-\Sigma _{33}\right)
\label{shear1}
\end{equation}
\begin{equation}
\Sigma _1=\frac 1{\sqrt{3}}\Sigma _{23},\quad \Sigma _3=\frac 1{\sqrt{3}%
}\Sigma _{12},\quad \tilde \Sigma _{13}= \frac 1{\sqrt{3}}\Sigma _{13}
\label{shear2}
\end{equation}
where for the sake of simplicity we can drop the tilde.

With these identifications the expansion-normalized variables $\left\{\Sigma
_+,\Sigma _-,\Sigma _1,\Sigma _3,\Sigma _{13},N_2,N_3,v_\alpha \right\}$
satisfy the set of evolution equations (A.11)-(A.13) and (A.29) plus the
algebraic constraint (A.18) given in \cite{Hewitt-Bridson-Wainwright}.
Consequently the state space is a compact subset $\mathcal{D}\subset \mathbf{%
R}^7$ (the compactness of $\mathcal{D}$ can be shown using the generalised
Friedmann equation and the inequality $\Omega>0$ for the energy density
parameter).

We conclude this section by noting that for each Bianchi type, the evolution
equations and the algebraic constraints can be determined by
specialising them in the standard way:

\begin{center}
\begin{tabular}{|l|l|l|l|}
\hline
\textbf{Type} & \textbf{N}$_2$ & \textbf{N}$_3$ & \textbf{Restrictions} \\ 
\hline
IV & $0$ & $N_3$ & $N_3>0$ \\ \hline
V & $0$ & $0$ & None \\ \hline
VI$_h$ & $N_2$ & $N_3$ & $N_2N_3<0$ \\ \hline
VII$_h$ & $N_2$ & $N_3$ & $N_2N_3>0$ \\ \hline
VI$_{-1/9}$ & $N_2$ & $N_3$ & $N_2N_3<0$ \\ \hline
\end{tabular}
\end{center}

\section{Solution of the symmetry equations}

\setcounter{equation}{0} In this section we will give the general solution of the
similarity equations $\mathcal{L}_{\mathbf{H}}g_{ab}=2\psi g_{ab}$ where $%
\mathbf{H}=H\mathbf{n}+A^\alpha \mathbf{X}_\alpha $ is the generator of the
one-parameter group of homotheties. Using the fact that $[\mathbf{H},\mathbf{%
X}_\alpha]=C_{4\alpha}^{\beta}\mathbf{X}_\beta$ and $[\mathbf{n},\mathbf{X}%
_\alpha]=0$ it follows $H=H(t)$ and $A=A(x^\alpha)$ where $H(t),A(x^\alpha)$
are smooth functions of the spacetime manifold \cite{Apostol9}. We note that the exact 
form of the self-similar SH metrics is \emph{independent} from the physical assumption we made 
for the matter content of the spacetime. However if we employ a perfect fluid source for the gravitational field, 
the well known relation $\mathcal{L}_{\mathbf{H}}R_{ab}=\mathcal{L}_{\mathbf{%
H}}T_{ab}=0$ implies \cite{Carot-Sintes} that the fluid velocity is mapped conformally by the HVF i.e. 
$\mathcal{L}_{\mathbf{H}}u_a=\psi u_a$.

In what follows the coordinate forms of the KVFs, the invariant basis and
its dual are those which appear in \cite{Ryan-Shepley1}. We recall that in
types VI$_h$ and VII$_h$ the group parameters $h_R$ and $h$ are related
according to $h=-\frac{\left( h_R+1\right) ^2}{\left( 1-h_R\right) ^2}$ and $%
h=\frac{h_R^2}{4-h_R^2} $ respectively.\vspace{0.5cm}

\noindent\underline{\textbf{Type III}}

In contrast with the type VI$_{-1/9}$ models ($h=-1/9\Leftrightarrow h_R=-2$
or $h_R=-1/2$) in which its ''exceptional'' dynamical behaviour do not pass
to geometry (at least regarding the determination of the corresponding
self-similar metrics), type III models ($h=-1\Leftrightarrow h_R=0$) cannot
be treated simultaneously within the class of type VI$_h$ metrics due to the
''singular'' property of the value $h_R=0$ in Jacobi identities.

In local coordinates $\{t,x,y,z\}$ the KVFs $\{\mathbf{X}_\alpha \}$ and the
canonical 1-forms $\{\mathbf{\omega }^\alpha \}$ are:

\begin{equation}
\mathbf{X}_1=\partial _y,\qquad \mathbf{X}_2=\partial _z,\qquad \mathbf{X}%
_3=\partial _x+y\partial _y  \label{typeIII.1}
\end{equation}
\begin{equation}
\mathbf{\omega }^1=e^{-x}dy,\qquad \mathbf{\omega }^2=dz,\qquad \mathbf{%
\omega }^3=dx.  \label{typeIII.2}
\end{equation}
The non-vanishing structure constants $C_{13}^1=1$ and the Jacobi identities
of the homothetic Lie algebra $\mathcal{H}_4$ imply that the remaining
non-vanishing structure constants $C_{\beta 4}^\alpha $ are either $%
C_{24}^2=b$ or $C_{34}^2=c$. We consider subcases (in order to pertain the
generality of the results we avoid to set some group constants equal to 1): 
\vspace{0.5cm}

\noindent \underline{Case A$_1$}

\noindent \textbf{HVF}

\begin{equation}
\mathbf{H}=\psi t\partial _t+D_1\partial _x+bz\partial _z  \label{typeIII.3}
\end{equation}
\textbf{Fluid velocity}

\begin{equation}
\Delta _1=v_1t^{(D_1+\psi )/\psi },\qquad \Delta _2=v_2t^{(\psi -b)/\psi
},\qquad \Delta _3=v_3t  \label{typeIII.4}
\end{equation}
\textbf{Metric}

\begin{equation}
g_{\alpha \beta }=\left( 
\begin{array}{ccc}
c_{11}t^{2(D_1+\psi )/\psi } & c_{12}t^{(D_1+2\psi -b)/\psi } & 
c_{13}t^{(2\psi +D_1)/\psi } \\ 
c_{12}t^{(D_1+2\psi -b)/\psi } & c_{22}t^{2(\psi -b)/\psi } & 
c_{23}t^{(2\psi -b)/\psi } \\ 
c_{13}t^{(2\psi +D_1)/\psi } & c_{23}t^{(2\psi -b)/\psi } & c_{33}t^2
\end{array}
\right) .  \label{typeIII.5}
\end{equation}
\vspace{0.5cm}

\noindent \underline{Case A$_2$}

\noindent \textbf{HVF}

\begin{equation}
\mathbf{H}=\psi t\partial _t+D_1\partial _x+D_2e^x\partial _y+cx\partial _z
\label{typeIII.6}
\end{equation}
\textbf{Fluid velocity}

\begin{equation}
\Delta _1=v_1t^{(D_1+\psi )/\psi },\qquad \Delta _2=v_2t,\qquad \Delta _3=-%
\frac{D_2v_1t^{(D_1+\psi )/\psi }}{D_1}-\frac{cv_2t\ln t}\psi +v_3t
\label{typeIII.7}
\end{equation}
\textbf{Metric}

\begin{equation}
g_{\alpha \beta }=\left( 
\begin{array}{ccc}
c_{11}t^{2(\psi +D_1)/\psi } & g_{21} & g_{31} \\ 
c_{12}t^{(2\psi +D_1)/\psi } & c_{22}t^2 & g_{32} \\ 
t^{\left( 2\psi +D_1\right) /\psi }\left( c_{13}-\frac{c_{12}c}\psi \ln
t\right) -\frac{D_2c_{11}t^{2(\psi +D_1)/\psi }}{D_1} & g_{32} & g_{33}
\end{array}
\right)  \label{typeIII.8}
\end{equation}
where 
\begin{equation}
g_{32}=t^2\left( c_{23}-\frac{c_{22}c\ln t}\psi \right) -\frac{%
c_{12}D_2t^{(2\psi +D_1)/\psi }}{D_1}  \label{typeIII.9}
\end{equation}
\begin{eqnarray}
g_{33} &=&\frac{D_2^2c_{11}}{D_1^2}t^{2(D_1+\psi )/\psi }+t^{(D_1+2\psi
)/\psi }\left( \frac{2D_2c_{12}c\ln t}{D_1\psi }-\frac{2D_2c_{13}}{D_1}%
\right) +  \nonumber \\
&&+\left[ \frac{c^2c_{22}(\ln t)^2}{\psi ^2}-\frac{2c_{23}c\ln t}\psi
+c_{33}\right] t^2.  \label{typeIII.10}
\end{eqnarray}
\vspace{0.5cm}

\noindent \underline{Case A$_3$}\vspace{0.5cm}

\noindent \textbf{HVF}

\begin{equation}
\mathbf{H}=\psi t\partial _t+D_2e^x\partial _y+cx\partial _z
\label{typeIII.11}
\end{equation}
\textbf{Fluid velocity}

\begin{equation}
\Delta _1=v_1t,\quad \Delta _2=v_2t,\quad \Delta _3=t\left( v_3-\frac{%
D_2v_1+cv_2}\psi \ln t\right)  \label{typeIII.12}
\end{equation}

\textbf{Metric}

\begin{equation}
g_{\alpha \beta }=\left( 
\begin{array}{ccc}
c_{11}t^2 & g_{21} & g_{31} \\ 
c_{12}t^2 & c_{22}t^2 & g_{32} \\ 
t^2\left( c_{13}-\frac{D_2c_{11}+c_{12}c}\psi \ln t\right) & t^2\left(
c_{23}-\frac{D_2c_{12}+c_{22}c}\psi \ln t\right) & g_{33}
\end{array}
\right)  \label{typeIII.12a}
\end{equation}
where $g_{33}=t^2\left[ c_{33}+\frac{D_2^2c_{11}+2D_2c_{12}c+c^2c_{22}}{\psi
^2}\left( \ln t\right) ^2-\frac{2\left( D_2c_{13}+c_{23}c\right) \ln t}\psi
\right] $.\vspace{0.5cm}
\clearpage
\noindent \underline{\textbf{Type IV}}

In type IV models, the KVFs and the dual basis have the following coordinate
forms:

\begin{equation}
\mathbf{X}_1=\partial _y,\qquad \mathbf{X}_2=\partial _z,\qquad \mathbf{X}%
_3=\partial _x+\left( y+z\right) \partial _y+z\partial _z  \label{typeIV.1}
\end{equation}
\begin{equation}
\mathbf{\omega }^1=e^{-x}dy-xe^{-x}dz,\qquad \mathbf{\omega }%
^2=e^{-x}dz,\qquad \mathbf{\omega }^3=dx.  \label{typeIV.2}
\end{equation}
Since the non-vanishing structure constants are $C_{13}^1=C_{23}^1=C_{23}^2=1
$ a similar analysis shows that the remaining non-vanishing structure
constants $C_{\beta 4}^\alpha $ of $\mathcal{H}_4$ are $C_{14}^1=a,\quad
C_{24}^2=a$. Therefore we have the following results:\ \vspace{0.5cm}

\noindent \textbf{HVF}

\begin{equation}
\mathbf{H}=\psi t\partial _t+D_1\partial _x+ay\partial _y+az\partial _z
\label{typeIV.4}
\end{equation}
\textbf{Fluid velocity}

\begin{equation}
\Delta _1=v_1t^{(D_1+\psi -a)/\psi },\qquad \Delta _2=t^{(D_1+\psi -a)/\psi
}\left( v_2+\frac{D_1v_1\ln t}\psi \right) ,\qquad \Delta _3=v_3t
\label{typeIV.5}
\end{equation}
\textbf{Metric}

\begin{equation}
g_{\alpha \beta }=\left( 
\begin{array}{ccc}
c_{11}t^{2(D_1+\psi -a)/\psi } & g_{21} & g_{31} \\ 
t^{2(D_1+\psi -a)/\psi }\left( c_{12}+\frac{D_1c_{11}\ln t}\psi \right) & 
g_{22} & g_{32} \\ 
c_{13}t^{(D_1+2\psi -a)/\psi } & t^{(D_1+2\psi -a)/\psi }\left( c_{23}+\frac{%
D_1c_{13}\ln t}\psi \right) & c_{33}t^2
\end{array}
\right)  \label{typeIV.6}
\end{equation}
where 
\begin{equation}
g_{22}=t^{2(D_1+\psi -a)/\psi }\left[ c_{22}+\frac{2D_1c_{12}\ln t}\psi +%
\frac{D_1^2c_{11}\left( \ln t\right) ^2}{\psi ^2}\right] .  \label{typeIV.7}
\end{equation}
\vspace{0.5cm} 
\noindent \underline{\textbf{Type V}}

In this case the KVFs $\{\mathbf{X}_\alpha \}$ and the canonical 1-forms $\{%
\mathbf{\omega }^\alpha \}$ are:

\begin{equation}
\mathbf{X}_1=\partial _y,\qquad \mathbf{X}_2=\partial _z,\qquad \mathbf{X}%
_3=\partial _x+y\partial _y+z\partial _z.  \label{typeV.1}
\end{equation}
\begin{equation}
\mathbf{\omega }^1=e^{-x}dy,\qquad \mathbf{\omega }^2=e^{-x}dz,\qquad 
\mathbf{\omega }^3=dx  \label{typeV.2}
\end{equation}
It turns out that the structure of the homothetic algebra is $%
C_{13}^1=C_{23}^2=1,\quad C_{14}^2=b,\quad C_{24}^1=a$. The analysis of the
symmetry equations suggests that we must again consider subcases according
to the vanishing of the parameter $a$: 

\vspace{0.5cm} \noindent \underline{Case A$_1$ ($a\neq 0$)}

\noindent \textbf{HVF}

\begin{equation}
\mathbf{H}=\psi t\partial _t+D_1\partial _x+bz\partial _y+ay\partial _z
\label{typeV.4}
\end{equation}
\textbf{Fluid velocity}

\begin{equation}
\Delta _1=v_1e^{\frac{(D_1-\sqrt{ab}+\psi )\ln t}\psi }+v_2e^{\frac{(D_1+%
\sqrt{ab}+\psi )\ln t}\psi },  \label{typeV.5}
\end{equation}
\begin{equation}
\Delta _2=-\frac{\psi t(\Delta _1)_{,t}-(D_1+\psi )\Delta _1}a,\qquad \Delta
_3=v_3t  \label{typeV.5a}
\end{equation}
\textbf{Metric} 
\begin{equation}
g_{11}=c_{12}e^{\frac{2(D_1-\sqrt{ab}+\psi )\ln t}\psi }+c_{21}e^{\frac{%
2(D_1+\sqrt{ab}+\psi )\ln t}\psi }+c_{23}t^{\frac{2(D_1+\psi )}\psi }
\label{typeV.6}
\end{equation}
\begin{equation}
g_{12}=-\frac{\psi t(g_{11})_{,t}-2(D_1+\psi )g_{11}}{2a},\quad
g_{13}=c_{13}e^{\frac{(D_1-\sqrt{ab}+2\psi )\ln t}\psi }+c_{31}e^{\frac{(D_1+%
\sqrt{ab}+2\psi )\ln t}\psi }  \label{typeV.7}
\end{equation}
\begin{equation}
g_{22}=\frac{\psi ^2t^2(g_{11})_{,tt}-\psi t(4D_1+3\psi
)(g_{11})_{,t}+2(2D_1^2+4D_1\psi -ab+2\psi ^2)g_{11}}{2a^2}  \label{typeV.9}
\end{equation}
\begin{equation}
g_{23}=-\frac{\psi t(g_{13})_{,t}-(D_1+2\psi )g_{13}}a,\quad g_{33}=c_{33}t^2
\label{typeV.10}
\end{equation}
\vspace{0.5cm}

\noindent \underline{Case A$_2$}\vspace{0.5cm}

\noindent \textbf{HVF}

\begin{equation}
\mathbf{H}=\psi t\partial _t+D_1\partial _x+bz\partial _y  \label{typeV.12}
\end{equation}
\textbf{Fluid velocity}

\begin{equation}
\Delta _1=v_1t^{\frac{D_1+\psi }\psi },\quad \Delta _2=t^{\frac{D_1+\psi }%
\psi }\left( v_2-\frac{bv_1\ln t}\psi \right) ,\quad \Delta _3=v_3t
\label{typeV.13}
\end{equation}

\textbf{Metric}

\begin{equation}
g_{\alpha \beta }=\left( 
\begin{array}{ccc}
c_{11}t^{\frac{2(D_1+\psi )}\psi } & g_{21} & g_{31} \\ 
t^{\frac{2(D_1+\psi )}\psi }\left( c_{12}-\frac{bc_{11}\ln t}\psi \right) & 
t^{\frac{2(D_1+\psi )}\psi }\left[ c_{22}-\frac{2bc_{12}\ln t}\psi +\frac{%
b^2c_{11}(\ln t)^2}{\psi ^2}\right] & g_{32} \\ 
c_{13}t^{\frac{D_1+2\psi }\psi } & t^{\frac{D_1+2\psi }\psi }\left( c_{23}-%
\frac{bc_{13}\ln t}\psi \right) & c_{33}t^2
\end{array}
\right) .  \label{typeV.14}
\end{equation}
\vspace{0.5cm}

\noindent \underline{\textbf{Type VI}$_{h_R}$}

The KVFs $\{\mathbf{X}_\alpha \}$ and the canonical 1-forms $\{\mathbf{%
\omega }^\alpha \}$ are:

\begin{equation}
\mathbf{X}_1=\partial _y,\qquad \mathbf{X}_2=\partial _z,\qquad \mathbf{X}%
_3=\partial _x+y\partial _y+h_Rz\partial _z.  \label{typeVI.1}
\end{equation}
\begin{equation}
\mathbf{\omega }^1=e^{-x}dy,\qquad \mathbf{\omega }^2=e^{-h_Rx}dz,\qquad 
\mathbf{\omega }^3=dx  \label{typeVI.2}
\end{equation}
where $h_R\neq 0$ and $C_{13}^1=h_R^{-1}C_{23}^2=1$. Hence the remaining
non-vanishing structure constants $C_{\beta 4}^\alpha $ are $%
C_{14}^1=a,\quad C_{24}^2=b$. Using these results we get: \vspace{0.5cm}

\noindent \textbf{HVF}

\begin{equation}
\mathbf{H}=\psi t\partial _t+D_1\partial _x+ay\partial _y+bz\partial _z
\label{typeVI.4}
\end{equation}
\textbf{Fluid velocity}

\begin{equation}
\Delta _1=v_1t^{(D_1+\psi -a)/\psi },\qquad \Delta _2=v_2t^{(D_1h_R+\psi
-b)/\psi },\qquad \Delta _3=v_3t  \label{typeVI.5}
\end{equation}
\textbf{Metric}

\begin{equation}
g_{\alpha \beta }=\left( 
\begin{array}{ccc}
c_{11}t^{2(D_1+\psi -a)/\psi } & g_{21} & g_{31} \\ 
c_{12}t^{\left[ D_1\left( h_R+1\right) +2\psi -a-b\right] /\psi } & 
c_{22}t^{2(D_1h_R+\psi -b)/\psi } & g_{32} \\ 
c_{13}t^{(D_1+2\psi -a)/\psi } & c_{23}t^{(D_1h_R+2\psi -b)/\psi } & 
c_{33}t^2
\end{array}
\right) .  \label{typeVI.6}
\end{equation}
\vspace{0.5cm}

\noindent \underline{\textbf{Type VII}$_{h_R}$}

Finally in type VII$_{h_R}$ models we have:

\begin{equation}
\mathbf{X}_1=\partial _y,\qquad \mathbf{X}_2=\partial _z,\qquad \mathbf{X}%
_3=\partial _x-z\partial _y+\left( y+h_Rz\right) \partial _z.
\label{typeVII.1}
\end{equation}
\begin{equation}
\mathbf{\omega }^1=\left( A_1-\frac{h_R}2A_2\right) dy-A_2dz,\quad \mathbf{%
\omega }^2=A_2dy+\left( A_1+\frac{h_R}2A_2\right) dz,\quad \mathbf{\omega }%
^3=dx  \label{typeVII.2}
\end{equation}
where: 
\begin{equation}
A_1=e^{-\frac{h_R}2x}\cos wx,\quad A_2=-\frac 1we^{-\frac{h_R}2x}\sin
wx,\quad w=\sqrt{\frac{4-h_R^2}4.}  \label{typeVII.3}
\end{equation}
\vspace{0.5cm}

\noindent The non-vanishing structure constants are ($h_R^2<4$) $%
C_{13}^2=-C_{23}^1=h_R^{-1}C_{23}^2=1$ leading to the following
expressions:\ 
\begin{equation}
C_{14}^1=a,\quad C_{24}^2=a.  \label{typeVII.5}
\end{equation}
\vspace{0.5cm}

\noindent \textbf{HVF}

\begin{equation}
\mathbf{H}=\psi t\partial _t+D_1\partial _x+ay\partial _y+az\partial _z
\label{typeVII.6}
\end{equation}
\textbf{Fluid velocity}

\begin{equation}
\Delta _1=t^{p_1/2\psi }\left[ v_{12}\cos \left( \frac{D_1w}\psi \ln
t\right) +v_{21}\sin \left( \frac{D_1w}\psi \ln t\right) \right] ,
\label{typeVII.7}
\end{equation}
\begin{equation}
\Delta _2=\frac{\psi t\left( \Delta _1\right) _{,t}+\left( a-\psi \right)
\Delta _1}{D_1},\qquad \Delta _3=v_3t.  \label{typeVII.8}
\end{equation}
\textbf{Metric}

\begin{equation}
g_{11}=-\frac{t^{p_1/\psi }\left\{ 16D_1E_1w^3p_1^2\cos \left( \frac{%
2D_1w\ln t}\psi \right) +4E_2w^2p_1\sin \left( \frac{2D_1w\ln t}\psi \right)
\right\} }{64D_1^3w^5p_1^2}  \label{typeVII.9}
\end{equation}
\begin{equation}
g_{12}=\frac{\psi t\left( g_{11}\right) _{,t}+2(a-\psi )g_{11}}{2D_1}
\label{typeVII.10}
\end{equation}
\begin{equation}
g_{13}=t^{(p_1+2\psi )/2\psi }\left[ c_{13}\cos \left( \frac{D_1w\ln t}\psi
\right) +c_{31}\sin \left( \frac{D_1w\ln t}\psi \right) \right] 
\label{typeVII.11}
\end{equation}
\begin{equation}
g_{22}=\frac{\psi ^2t^2\left( g_{11}\right) _{,tt}-\psi t\left(
g_{11}\right) _{,t}(p_1+\psi -2a)+2\left[ D_1^2+(\psi -a)p_1\right] g_{11}}{%
2D_1^2}  \label{typeVII.12}
\end{equation}
\begin{equation}
g_{23}=\frac{\psi t\left( g_{13}\right) _{,t}+(a-2\psi )g_{13}}{D_1},\quad
g_{33}=c_{33}t^2  \label{typeVII.14}
\end{equation}
where $E_1,E_2$ are constants and we have set $p_1=D_1h_R+2\left( \psi
-a\right) $. 

\section{Exact Solutions}

\setcounter{equation}{0} The well known feature of self-similarity to reduce
the FE to a purely algebraic form, makes the determination of vacuum and
perfect fluid models straightforward. Following the methodology of paper I
and for illustration purposes, we reproduce all the known vacuum and
non-tilted perfect fluid solutions. In addition we find the \emph{general}
tilted perfect fluid model for the exceptional class VI$_{-1/9}$. For
completeness we also give, for each Bianchi type, the extreme\footnote{%
We recall that an extreme tilted equilibrium point satisfies $v^\alpha v_\alpha=1$.}
tilted equilibrium points of their state space. \vspace{0.5cm}

\noindent \underline{\textbf{Vacuum plane wave models of type III}}

This exact vacuum solution can be found from the case A$_1$ of section 2 and
has been given in \cite{Hsu-Wainwright}. The HVF is $\mathbf{H}=\psi
t\partial _t+D_1\partial _x+bz\partial _z$ and the constants $c_{\alpha
\beta },\psi ,b$ are:

\begin{equation}
c_{13} =c_{23}=0, \quad  c_{11} =\frac{c_{12}^2(D_1-b)}{2c_{22}D_1},\quad c_{33}=\frac{b^2}{\left( D_1+b\right) ^2},\quad \psi =b 
\label{vacIII.1} 
\end{equation}
where without loss of generality we can set $b=1$.

We note that the above spacetime is algebraically special (Petrov type N)
since it admits the \emph{gradient null KVF} $l^a=e^{-bx/(D_1+b)}\left(
\partial _t+\frac{D_1+b}{bt}\partial _x\right) $ which essentially
represents the repeated principal null direction of the Weyl tensor. \vspace{%
0.5cm}

\noindent \underline{\textbf{Non-tilted perfect fluid models of type III}}

It is a special case of the non-tilted perfect fluid solution found by
Collins \cite{Collins2} for $h=-1$. The HVF is $\mathbf{H}=\psi t\partial
_t+bz\partial _z$ and the constants $c_{\alpha \beta },\psi ,b$ are given by:

\begin{eqnarray}
c_{13} &=&c_{23}=c_{12}=0,\quad c_{11}=c_{22}=1  \nonumber \\
&&  \label{nontiltIII.1} \\
c_{33} &=&\frac{\gamma ^2}{\left( 2-\gamma \right) \left( 3\gamma -2\right) }%
,\quad \mu =\frac{4\left( 1-\gamma \right) }{\gamma ^2t^2},\quad \psi =\frac{%
b\gamma }{3\gamma -2}.  \nonumber
\end{eqnarray}
\vspace{0.5cm}

\noindent \underline{\textbf{Tilted perfect fluid models of type III}}

Using the results of section 2 we can show that there is no physically
acceptable tilted perfect solution of type III. We note that this conclusion
can be also confirmed from the set of the tilted equilibrium points of the
dynamical state space given in \cite{Apostol15}. \vspace{0.5cm}

\noindent \underline{\textbf{Vacuum plane wave models of type IV}}

From the self-similar metrics (\ref{typeIV.6}) we can find the general type
IV vacuum solution, first given in \cite{Harvey-Tsoubelis}. In this case the
HVF is $\mathbf{H}=\psi t\partial _t+D_1\partial _x+ay\partial _y+az\partial
_z$ or, by performing a suitable change of the basis of the homothetic Lie
algebra, $\mathbf{H}=\psi t\partial _t+\left( D_1-a\right) \partial
_x-az\partial _y$. In addition the constants $c_{\alpha \beta },\psi $ are 
given by:

\begin{eqnarray}
c_{13} =c_{23}=0,\quad c_{22} =\frac{D_1(c_{11}^2-4c_{12}^2)+4ac_{12}^2}{4c_{11}\left(a- D_1\right) },\quad c_{33}=\frac{a^2}{D_1^2},\quad \psi =a.  
\label{vacIV.1}
\end{eqnarray}
It can be verified that the above one parameter family of models is also
algebraically special and admits the \emph{gradient null} KVF $%
l^a=e^{-ax/D_1}\left( \partial _t+\frac{D_1}{at}\partial _x\right) $.\vspace{%
0.5cm}

\noindent \underline{\textbf{Non-tilted and tilted perfect fluid models of
type IV}}

It is found that no orthogonal or tilted perfect fluid self-similar models
of type IV exist. Nevertheless there exists the following extreme tilted
equilibrium point \cite{Hervik-Hoogen-Coley}: 
\[
N_2=0,\quad N_3=N_3,\quad v^\alpha v_\alpha =1, 
\]
\[
\Sigma ^2=\frac{N_3^2+3q^2}{12},\quad \Sigma _3=\Sigma _{13}=\Sigma
_{-}=0,\quad \Sigma _{+}=-\frac q2, 
\]
\[
\Sigma _1=\frac{\sqrt{3}N_3}6,\quad v_1=1,\quad A_1=\frac{2-q}2 
\]
\[
\Omega =\frac{3q(2-q)-N_3^2}6,\qquad 0<\gamma <2. 
\]
\vspace{0.4cm}
\noindent \underline{\textbf{Models of type V}}

The analysis shown that no self-similar vacuum or perfect fluid models exist
for type V. This result can also be proved by using the set of evolution
equations given in \cite{Hewitt-Bridson-Wainwright} (see also \cite
{Apostol15}). However there exists the following extreme tilted equilibrium
point \cite{Hewitt-Wainwright2,Coley-Hervik}: 
\[
N_2=N_3=0,\quad v^\alpha v_\alpha =1, 
\]
\[
\Sigma ^2=\left( A_1-1\right) ^2,\quad \Sigma _3=\Sigma _{13}=\Sigma
_1=\Sigma _{-}=0,\quad \Sigma _{+}=-\frac q2, 
\]
\[
v_1=1,\quad A_1=\frac{2-q}2,\quad \Omega =2A_1(1-A_1),\quad 0<\gamma <2. 
\]
 
\vspace{0.8cm}

\noindent \underline{\textbf{Vacuum plane wave models of type VI}$_{h_R}$}

In this case the HVF is $\mathbf{H}=\psi t\partial _t+D_1\partial
_x+ay\partial _y+bz\partial _z$ and the constants $c_{\alpha \beta },\psi $
are given by \cite{Hsu-Wainwright}: 
\begin{eqnarray}
c_{13} &=&c_{23}=0,\quad c_{11}=\frac{c_{12}^2\left(
h_R+1\right) \left[ D_1\left( h_R+1\right) -a-b\right] }{2c_{22}\left[ D_1\left(
h_R^2+1\right) -a-bh_R\right] }  \nonumber \\
&&  \label{vacVIh.1} \\
c_{33} &=&\frac{(ah_R-b)^2}{\left[ D_1\left( h_R-1\right) +a-b
\right] ^2},\quad \psi =\frac{ah_R-b}{h_R-1}.  \nonumber
\end{eqnarray}
\vspace{0.5cm}

\noindent The \emph{gradient null KVF} is $l^a=e^{-\frac{ah_R-b}{D_1\left(
h_R-1\right) +a-b }x}\left[ \partial _t+\frac{D_1\left(
h_R-1\right) +a-b}{\left( ah_R-b\right) t}\partial _x\right] $. 

As we have mentioned in section 2, in order to determine the
corresponding self-similar solution for the exceptional type VI$_{-1/9}$
models we can use the geometric results (\ref{typeVI.4})-(\ref{typeVI.6})
and solve the FE by setting $h_R=-2$ or $h_R=-1/2$ (for these values of the
group parameter the $01$ or $02$ component of (\ref{prel4}) vanish and essentially
represents the ''exceptional'' behaviour of the VI$_{-1/9}$ models in terms
of the FE). The resulting solution has been first given by Robinson and
Trautman (RT) \cite{Robinson-Trautman} (see also \cite{Hsu-Wainwright}) and
has the following form ($c_{13}\neq 0$):

\begin{equation}
c_{12} =c_{23}=0,\quad c_{33}=\frac{75}8,\quad c_{11}=\frac{24c_{13}^2}{125},\quad D_1 =-\frac b3,\quad \psi =\frac{5b}6,\quad a=\frac b3.
 \label{vacVIh.2}\
\end{equation}
\vspace{0.5cm}

\noindent \underline{\textbf{Non-tilted perfect fluid models of type VI}$%
_{h_R}$}

The family of non-tilted perfect fluid models has been found by Collins \cite
{Collins2}. The HVF is given by equation (\ref{typeVI.4}) and the constants $%
c_{\alpha \beta },\psi ,a,b$ are:

\begin{eqnarray}
c_{13} &=&c_{23}=c_{12}=0,\quad c_{11}=c_{22}=1,\quad c_{33}=\frac{\gamma
^2\left( h_R-1\right) ^2}{\left( 2-\gamma \right) \left( 3\gamma -2\right) }
\nonumber \\
&&  \nonumber \\
\mu &=&-\frac{4\left[ h_R^2\left( \gamma -1\right) +h_R\gamma +\gamma
-1\right] }{\gamma ^2\left( h_R-1\right) ^2t^2},\quad \psi =\frac{\gamma \left(
D_1-a\right) \left( h_R-1\right) }{h_R\left( 2-3\gamma \right) }
\label{nontiltVIh.1} \\
&&  \nonumber \\
b &=&\frac{D_1\left( h_R^2+1\right) -a}{h_R}.  \nonumber
\end{eqnarray}
The exceptional type VI$_{-1/9}$ non-tilted perfect fluid solution can be
found similarily by setting $h_R=-2$ (or $h_R=-1/2$) and has been given in \cite{Wainwright}: 
\begin{eqnarray}
c_{12} &=&c_{23}=0,\quad c_{33}=\frac{75}8,\quad \psi =\frac{5b}2,\quad
D_1=0,\quad a=2b  \nonumber \\
&&  \label{nontiltVIh.2} \\
\mu &=&\frac{9(125c_{11}-24c_{13}^2)}{25(75c_{11}-8c_{13}^2)t^2},\quad \gamma =%
\frac{10}9.  \nonumber
\end{eqnarray}
\clearpage
\noindent \underline{\textbf{Tilted perfect fluid models of type VI}$_{h_R}$}

It is convenient to employ the constants $p_1,p_2$: 
\begin{equation}
D_1=a+p_1\psi -2\psi ,\quad a=bh_R-\psi \left( p_1+p_2-4\right) .
\label{tiltVIh.1}
\end{equation}
The frame components of the metric and the fluid velocity become: 
\begin{equation}
g_{\alpha \beta }=\left( 
\begin{array}{ccc}
c_{11}t^{2(p_1-1)} & g_{21} & g_{31} \\ 
c_{12}t^{[b(h_R^2-1)+h_R\psi (2-p_2)+p_1\psi ]/\psi } & 
c_{22}t^{2[b(h_R^2-1)+h_R\psi (2-p_2)+\psi ]/\psi } & g_{32} \\ 
c_{13}t^{p_1} & c_{23}t^{[b(h_R^2-1)+h_R\psi (2-p_2)+2\psi ]/\psi } & 
c_{33}t^2
\end{array}
\right)  \label{tiltVIh.3}
\end{equation}
\begin{equation}
\Delta _1=v_1t^{p_1-1,}\qquad \Delta _2=v_2t^{[b(h_R^2-1)+h_R\psi
(2-p_2)+\psi ]/\psi },\qquad \Delta _3=v_3t  \label{tiltVIh.4}
\end{equation}
whereas the HVF assumes the form:

\begin{equation}
\mathbf{H}=\psi t\partial _t+\left[ bh_R-\psi \left( p_2-2\right) \right]
\partial _x+\left[ bh_R-\psi (p_1+p_2-4)\right] y\partial _y+bz\partial _z.
\label{tiltVIh.5}
\end{equation}
For the case $h_R\neq -2$ (or $h_R\neq -1/2$) the general self-similar
solution has been found recently \cite{Apostol15}. Here we give the
corresponding \emph{general solution} for the type VI$_{-1/9}$ models.

We define the parameter $s$ according to: 
\begin{equation}
\gamma =\frac 2{2s+1}.  \label{tiltVIh.7}
\end{equation}
Then, the 00--conservation equation implies that the constant $p_1$ is given
by: 
\begin{equation}
p_1=\frac{2(6s+1)-p_2}4.  \label{tiltVIh.2}
\end{equation}
The various integration constants are:

\begin{equation}
c_{12}=c_{13}=v_1=0,\quad \psi =-\frac {2b}{p_2-2}  \label{tiltVIh.8}
\end{equation}
\begin{equation}
\begin{array}{l}
c_{22}=\frac{c_{23}^2(p_2-2)^2\left[ 2(3s-1)-p_2\right] \left[
-9p_2^2+4p_2(3-2s)+624s^2-208s+12\right] }{96s\left[
25p_2^2-20p_2(6s+1)+4(6s+1)^2\right] } \\ 
\\ 
c_{33}=\frac{48\left[ 3p_2^2+4p_2(8s-3)-4(12s^2+8s-3)\right] }{%
(p_2-2)^2\left[ 9p_2^2+4p_2(2s-3)-4(156s^2-52s+3\right] } \\ 
\\ 
\Gamma ^2=-\frac{s(5p_2-12s-2)\left[ 2(3s-1)-p_2\right] \left[
-9p_2^2+4p_2(3-2s)+624s^2-208s+12\right] }{2\left[ p_2+2(2s-1)\right] \left[
p_2^2(31s-6)+4p_2s(78s-17)-4(1476s^3-660s^2+101s-6)\right] } \\ 
\\ 
v_3=-\frac{36(p_2+4s-2)\left[ p_2+2(2s-1)\right] }{%
9p_2^3+2p_2^2(4s-15)-12p_2(52s^2-16s-1)+8(156s^2-52s+3)} \\ 
\\ 
v_2=\frac{3c_{23}\left[ p_2^2+4p_2(s-1)-4(2s-1)\right] }{4\left[
2(6s+1)-5p_2\right] }.
\end{array}
\label{tiltVIh.9}
\end{equation}
\vspace{0.5cm}

\noindent The constant $p_2$ is related with the ''state parameter'' $s$
according to:

\begin{equation}
\begin{array}{l}
p_2=\frac{2\left[ -6\sqrt{36s^4-204s^3+133s^2-22s+1}\left| 4s-1\right|
+s(336s^2-214s+19)\right] }{160s^2-91s+6}
\end{array}
\label{tiltVIh.10}
\end{equation}
and the energy density of the model is:

\begin{equation}
\begin{array}{l}
\tilde{\mu}=\frac{(2s+1)\left[
p_2^2(31s-6)+4p_2s(78s-17)-4(1476s^3-660s^2+101s-6\right] }{%
144s(p_2-12s+2)t^2}
\end{array}
\label{tiltVIh.11}
\end{equation}
In order to ensure the positivity of the energy density and the real values
of the constant $p_2$ (together with the overall signature of the metric)
this new exact solution is only defined when: 
\begin{equation}
s\in \left( \frac 16,\frac 14\right) \Leftrightarrow \gamma \in \left( \frac
43,\frac 32\right) .  \label{tiltVIh.12}
\end{equation}
\vspace{0.5cm}

\noindent It is interesting to note that, similarly with the types VI$_0$
and VI$_h$, the above \emph{rotating} cosmological model admits the \emph{%
hypersurface orthogonal} KVF $\mathbf{X}_2=\partial _z$ and belongs to the
subclass $n_\alpha ^\alpha =0$. In fact the present solution arises as an
equilibrium point of the tilted perfect fluid Bianchi type VI$_{-1/9}$
dynamical state space which we denote it as $\mathcal{E}$ and has the
following kinematical and dynamical quantities: \vspace{0.5cm}

\begin{tabular}{l}
$N_3=-\sqrt{-\frac{3\left[ q\left( 3\gamma -6\right) +7\gamma -6\right]
\left[ q^2\left( 4\gamma ^2-4\gamma -12\right) -2q\left( 3\gamma ^2-22\gamma
+12\right) -37\gamma ^2+48\gamma -12\right] }{16\gamma ^2\left[ q(\gamma
-2)+2(\gamma -1)\right] }}$ \\ 
$N_2=-N_3,\quad \Sigma _{-}=0,\quad \Sigma _{+}=-\frac q2,\quad v_2=-v_3$ \\ 
\\ 
$\Sigma _1=-\frac{\sqrt{3}\left[ q(5\gamma -12)+2(7\gamma -6)\right] }{%
6\gamma },\quad \Sigma _3=-\frac{\sqrt{6(2-\gamma )}[2q^2(\gamma
-3)+3q(3\gamma -4)+7\gamma -6]}{12\gamma \sqrt{(q+1)[q(\gamma -2)+2(\gamma
-1)]}}$ \\ 
\\ 
$\Sigma _{13}=\frac{\sqrt{6(2-\gamma )}[2q^3(\gamma -3)+q^2(11\gamma
-18)+2q(8\gamma -9)+7\gamma -6]}{12\gamma \sqrt{(q+1)^3[q(\gamma
-2)+2(\gamma -1)]}}$ \\ 
\\ 
$v_1=\frac{3(2q-3\gamma +2)}{2N_3\gamma },\quad v_3=-\frac{3\sqrt{2(2-\gamma
)(q+1)}(2q-3\gamma +2)}{8N_3\gamma \sqrt{q(\gamma -2)+2(\gamma -1)}},\quad
v^2=\frac{9(2q-3\gamma +2)^2[3q(\gamma -2)+7\gamma -6]}{16N_3^2\gamma
^2[q(\gamma -2)+2(\gamma -1)]}$ \\ 
\\ 
$\Omega =\frac{\left[ q\left( 3\gamma -6\right) +7\gamma -6\right] \left[
2q\left( \gamma -3\right) +7\gamma -6\right] \left[ 4q^2\left( 4-\gamma
\right) +2q\left( 8-15\gamma \right) +27\gamma ^2-26\gamma \right] }{%
12\gamma ^2(2q-3\gamma +2)[q(\gamma -2)+2(\gamma -1)]}$ \\ 
\\ 
$\Sigma ^2=\frac{[3q(\gamma -2)+7\gamma -6][4q^2(2\gamma ^2-8\gamma
+9)+4q(9\gamma ^2-29\gamma +18)+49\gamma ^2-84\gamma +36]}{12\gamma
^2[q(\gamma -2)+2(\gamma -1)]}$ \\ 
\\ 
$q=\frac{\gamma \sqrt{(\gamma -1)(73\gamma ^3-253\gamma ^2+240\gamma -36)}%
\left| 3\gamma -4\right| -2(\gamma -1)(33\gamma ^3-179\gamma ^2+294\gamma
-144)}{6(\gamma -1)(\gamma -2)(3\gamma ^2-19\gamma +24)}$%
\end{tabular}
\vspace{0.5cm}

\noindent We also note that there exists the following extreme tilted
equilibrium point \cite{Coley-Hervik}: \vspace{0.5cm} 
\[
N_2=N_2,\quad N_3=N_3,\quad v^\alpha v_\alpha =1, 
\]
\[
\Sigma ^2=\frac{3(N_3-N_2)^2-4(N_2N_3+6\sqrt{-N_2N_3}-9)}{36},\quad \Sigma
_3=\Sigma _{13}=\Sigma _{-}=0, 
\]
\[
\Sigma _1=\frac{\sqrt{3}(N_3-N_2)}6,\quad \Sigma _{+}=-\frac q2,\quad q=%
\frac{2(3-\sqrt{-N_2N_3})}3,\quad v_1=1, 
\]
\[
\Omega =-\frac{3(N_3-N_2)^2-4(N_2N_3+3\sqrt{-N_2N_3})}{18},\quad 0<\gamma
<2. 
\]
\vspace{0.5cm}
\noindent \underline{\textbf{Models of type VII}$_{h_R}$}

The question of the existence of tilted perfect type VII$_h$ models is
notoriously difficult to answer due to the complexity of the self-similar
metrics (\ref{typeVII.9})-(\ref{typeVII.14}). In fact the resulting FE are
difficult to handle analytically even to the case of expressing the known VII%
$_h$ vacuum solution (see e.g. \cite{Wainwright-Ellis} page 192 for an
elegant form of this solution and \cite{Barrow-Tsagas} for a complete study
of its stability properties against vorticity, shear and Weyl curvature
perturbations). Nevertheless, using the set of evolution equations we can
show that the assumption of the existence of a proper HVF is incompatible
with tilted perfect fluid models and that type VII$_h$ models possess the
following extreme tilted equilibrium point (this equilibrium point has been
also given in \cite{Hervik-Hoogen-Coley}): \vspace{0.5cm}

\[
N_2=N_2,\quad N_3=N_3,\quad v^\alpha v_\alpha =1, 
\]
\[
\Sigma ^2=\frac{(N_3-N_2)^2+12(hN_2N_3-2\sqrt{hN_2N_3}+1)}{12},\quad \Sigma
_3=\Sigma _{13}=\Sigma _{-}=0, 
\]
\[
\Sigma _1=\frac{\sqrt{3}(N_3-N_2)}6,\quad \Sigma _{+}=-\frac q2,\quad q=2(1-%
\sqrt{hN_2N_3}),\quad v_1=1, 
\]
\[
\Omega =-\frac{(N_3-N_2)^2+12(hN_2N_3-\sqrt{hN_2N_3})}6,\quad 0<\gamma <2. 
\]
\vspace{0.5cm}

\section{Discussion}

The main concern of papers I and II can be regarded having two main branches:

a) a geometric nature concerning the determination of all the SH geometries
restricted \emph{only} by the requirement of admitting a proper HVF and,

b) the physical implications of the general geometric results by finding, whenever
they exist, the corresponding vacuum or perfect fluid models.

The underlying importance of a) is that we have not incorporated a specific
form for the matter fields filled the spacetime therefore we can associate
the self-similar metrics with more general matter configurations with a view
to analyse their physical significance. On the other hand the well
established aspect of the SH equilibrium points as past or future attractors
for general vacuum or perfect fluid models necessitates the knowledge of all the self-similar
equilibrium points in order to gain deeper insight into their asymptotic
dynamics. This was accomplished in the series of articles I, II, \cite
{Apostol-Tsampa7,Apostol12,Apostol15} in which the \emph{complete set} of tilted
perfect fluid self-similar models have been found. For convenience we collect them in
table 1 together with the corresponding references.

As an immediate application of the geometric results of section 3, we
reproduced the known exact vacuum and non-tilted solutions for the Bianchi
types of class B and we have found the \emph{general} form of the
self-similar tilted type VI$_{-1/9}$ models. As we have seen in section 4
this \emph{new} exact solution arise as equilibrium point in the VI$_{-1/9}$
state space and shares many of the common properties of type VI$_0$ and VI$%
_h $ ($h\neq -1/9$) models and important differences as well. For example
the stability analysis of vacuum and non-tilted equilibrium points \cite
{Barrow-Hervik} has shown that the Collins VI$_{-1/9}$ solution is stable
for $\gamma \in \left( \frac 23,\frac{10}9\right) $. At the arc of
equilibrium points $\gamma =\frac{10}9$ \cite{Hewitt-Horwood-Wainwright}
there is an exchange of stability with the RT vacuum solution \cite
{Ellis-MacCallum,Robinson-Trautman} which is stable whenever $\gamma \in
\left( \frac{10}9,\frac 43\right) $. As a consequence we expected that
tilted models will be future dominated against the RT model when $\gamma
>\frac 43$. Indeed the equilibrium point $\mathcal{E}$ is found to have a
five dimensional stable manifold for $\gamma \in \left( \frac 43,\frac
32\right) $. However it can be shown that, at least for the case where $%
n_\alpha ^\alpha =0$, there exist always two unstable modes in the
restricted state space. One can made a step further and conjecture that for
the type VI$_h$ and VI$_{-1/9}$ models the values $\gamma =\frac{2\left( 3+%
\sqrt{-h}\right) }{5+3\sqrt{h}}$ and $\gamma =\frac 43$ respectively
represent line bifurcations similarly to the type VI$_0$ models with $\gamma
=\frac 65$. However no such bifurcations are found in both models.

\begin{table}[tbp]
\caption{This table contains all the self-similar tilted perfect fluid SH
models and the relevant range of the state parameter $\gamma$.}
\label{Tilted-self-similar models}
\begin{center}
\begin{tabular}{@{}lll}
\br \textbf{Type} & \textbf{Self-similar Model} & \textbf{Reference} \\ 
\mr I & $\qquad \qquad \qquad \nexists $ & \cite{King-Ellis} \\ 
II & $\qquad \qquad \gamma \in \left( \frac{10}7,2\right) $ & \cite{Hewitt}
\\ 
VI$_0$ & $\qquad \qquad \gamma \in \left[ \frac 65,\frac 32\right) $ & \cite
{Apostol12,Rosquist-Jantzen1} \\ 
VII$_0$ & $\qquad \qquad \qquad \nexists $ & \cite{Apostol9} \\ 
VIII & $\qquad \qquad \qquad \nexists $ & \cite{Apostol-Tsampa7} \\ 
IX & $\qquad \qquad \qquad \nexists $ & \cite{Apostol-Tsampa7} \\ 
III & $\qquad \qquad \qquad \nexists $ & \cite{Apostol15} \\ 
IV & $\qquad \qquad \qquad \nexists $ & paper II \\ 
V & $\qquad \qquad \qquad \nexists $ & paper II \\ 
&  &  \\ 
VI$_h$ & $
\begin{array}{l}
-1<h<-\frac 19\Rightarrow 1<\gamma <\frac{2\left( 3+\sqrt{-h}\right) }{5+3%
\sqrt{-h}} \\ 
\\ 
-\frac 19<h<0\Rightarrow \frac{2\left( 3+\sqrt{-h}\right) }{5+3\sqrt{-h}}%
<\gamma <\frac 32
\end{array}
$ & \cite{Apostol15} \\ 
&  &  \\ 
VII$_h$ & $\qquad \qquad \qquad \nexists $ & paper II \\ 
VI$_{-1/9}$ & $\qquad \qquad \gamma \in \left( \frac 43,\frac 32\right) $ & 
paper II \\ 
\br &  & 
\end{tabular}
\end{center}
\end{table}

Furthermore, for the remaining types, the mutual use of the metric and
orthonormal frame approaches guarantees that tilted perfect fluid models of
types III, IV, V and VII$_h$ cannot admit a proper HVF indicating that these
models either may not be asymptotically self-similar (a preliminary analysis
suggests that this behaviour occurs in type V models) or may be extreme
tilted at late times (see \cite{Hervik-Hoogen-Coley} for a thorough
discussion on the asymptotic dynamics of type IV models).

We conclude by noting that an interesting aspect of the analysis presented
in papers I and II is that vacuum and tilted perfect fluid SH self-similar
models exhibit further geometric constraints coming from the existence
of covariantly constant null vector fields or a hypersurface orthogonal KVF.
Therefore we believe that it will be interesting to study how the
assumptions of the self-similarity and a specific dynamic description lead
to extra restrictions on the geometry of the SH models.

\vskip 0.5cm

\centerline{\bf\large Acknowledgments} \vskip .3cm The author wishes to
acknowledge the financial support of Ministry of National Education and
Religious Affairs through the research program ''Pythagoras'', Grant No
70-03-7310. 

\end{document}